\documentclass[vecarrow]{svmult}
\usepackage{graphicx}
\usepackage{amssymb}
\usepackage{amsmath}
\usepackage{subeqnar}
\usepackage{epsfig}
\usepackage{citesort}

\begin{document}
\frontmatter
\mainmatter

\bibliographystyle{unsrt}

\title*{Vibrated granular media as experimentally realizable Granular Gases}
\toctitle{Vibrated granular media as experimentally realizable
\protect\newline Granular Gases}
\titlerunning{Vibrated granular media as experimentally realizable Granular Gases}
\author{Sean McNamara\inst{1,3} \and Eric Falcon\inst{2}}

\authorrunning{S. McNamara and E. Falcon}
\institute{$^1$ Centre Europ\'een de Calcul Atomique et Mol\'eculaire,\\
$^2$ Laboratoire de Physique, \'Ecole Normale Sup\'erieure de Lyon,\\ 
46, all\'ee d'Italie, 69007 Lyon, France\\
$^3$ current address: Institute for Computer Applications,\\
Universit\"at Stuttgart, 70569 Stuttgart, Germany}

\maketitle

\begin{abstract}
We report numerical simulations of strongly vibrated granular materials
designed to mimic recent experiments performed both in presence
\cite{Falcon:01} or 
absence \cite{Falcon:99} of gravity.  We show that a model with
impact velocity dependent restitution coefficient is necessary
to bring the simulations into agreement with experiments. We
measure the scaling exponents of the granular temperature, collision
frequency, impulse and pressure with the vibrating piston velocity. 
As the system
changes from a homogeneous gas state at low density to a clustered state at
high density, these exponents are all found to decrease continuously with
the particle number. In absence of gravity, a loose cluster appears near the upper wall, opposite the piston, and acts as a 
buffer for fastest particles leading to unexpected non-extensive scaling
exponents ; whereas in presence of gravity, the cluster bounces as a single
inelastic body. All these results differ significantly from classical
inelastic hard sphere kinetic theory and previous simulations, both based
on a constant restitution coefficient. 
\end{abstract}

\section{Introduction}

Granular gases \cite{GranularGases} often seem to be the exclusive domain of theory and
simulation.  For example, the ``freely cooling'' granular gas \cite{GranularGases} is impossible
to experimentally realize: even if an experimentalist somehow managed
to implement periodic boundaries in all directions, he would still have
to contend with energy and time scales that vary over many orders of
magnitude.  Another frequently studied example of granular gases is
the vibrated granular medium.  This case is much closer to experiments,
because it is relatively simple to put particles in a box and shake them,
or drive particles with a vibrating piston.  Indeed, there are
many experimental \cite{Warr:95,Yang:00,Luding:94a},
numerical \cite{Luding:94a,Luding:94b,McNamara:98},
and theoretical \cite{Warr:95,Huntley:98,Kumaran:98,Lee:95}
studies of this system.  However, numerous questions remain about
the link between experiments on one hand, and theory and simulations
on the other.  Most numerical and theoretical studies were not intended
to be compared with experiments.  Therefore, they have parameter
values far from the experimental ones, and none of them
predict even the most basic features of the experimental results.  
For example, numerical and theoretical studies
often assume that the vibration frequency is
very high so that the shaking wall can be replaced by a thermal
boundary.  The vibrated granular gas thus attains a steady state.
But very high
frequencies with an amplitude sufficient to fluidize the granular
materials are difficult to attain experimentally; 
the vibration amplitude in experiments is often
a significant fraction of the size of the container.  
In experiments, the pressure and granular temperature all vary strongly
over the period of one vibration cycle.  It has therefore been impossible
to compare the experimental and numerical results in a meaningful way.

In this paper, we bridge the gap between experiments and
numerics by presenting
simulations of strongly vibrated
granular materials designed to mimic recent experiments
\cite{Falcon:01,Falcon:99}. 
We present the first simulations which resemble
the experiments for a large range of parameters. 
We show that two parameters are especially important
for the agreement between experiment and simulation.
First of all, an
impact velocity dependent restitution coefficient is necessary
to bring the simulations into agreement with experiment. Most
previous studies consider only constant restitution coefficient 
\cite{Luding:94a,Luding:94b,McNamara:98}.
Secondly, it is important to explicitly consider the number
of particles $N$.  Studying only one value of $N$ or comparing
results obtained at different $N$ can lead to interpretive difficulties.

We measure the scaling exponents of the granular temperature, collision
frequency, impulse and pressure with the vibrating piston velocity, and find
results that differ significantly from classical
inelastic hard sphere kinetic theory and previous simulations.
Finally, we show that the system undergoes a smooth transition
from a homogeneous gas state at low density to a clustered state at
high density.  Depending on whether gravity is present or not, the
transition manifests itself in different ways: due to inelastic collisions (through the restitution coefficient properties), a nearly motionless loose cluster is observed in microgravity with non-extensive scaling properties, whereas a bouncing cluster appears under gravity.

\section{Description of the simulations}

\subsection{The variable coefficient of restitution}

The greatest difference between our simulations and the previous numerical
studies of vibrated granular media
\cite{Luding:94a,Luding:94b,McNamara:98} is that we use a restitution coefficient that depends on impact
velocity.  The restitution coefficient $r$ is the ratio between the relative normal velocities before and
after impact.  In all previous simulations of strongly vibrated
granular media, the
coefficient of restitution is considered to be constant and lower
than 1.  However, since a century, it has been shown from impact
experiments that $r$ is a function of the impact velocity $v$
\cite{Raman:1918,Johnson:85,Labous:97,Kuwabara:87,Falcon:98}. Indeed, for
metallic particles, when $v$ is large ($v \gtrsim 5$ m/s \cite{Johnson:85}),
the colliding particles deform fully plastically and $r \propto v^{-1/4}$
\cite{Raman:1918,Johnson:85,Labous:97}. When $v\lesssim 0.1$ m/s
\cite{Johnson:85}, the deformations are elastic with mainly viscoelastic
dissipation, and $1-r \propto v^{1/5}$
\cite{Labous:97,Kuwabara:87,Falcon:98,Hertzsch:95}. Such velocity-dependent
restitution coefficient models have recently shown to be important in
numerical
\cite{Brilliantov:00,Saluena:99,Bizon:98,Poschel:01,Goldman:98,Salo:88} and
experimental \cite{Falcon:98,Bridges:84} studies. Applications include: 
granular fluidlike properties (convection \cite{Saluena:99}, surface waves
\cite{Bizon:98}), collective collisional processes (energy transmission
\cite{Poschel:01}, absence of collapse \cite{Falcon:98,Goldman:98}) and
planetary rings \cite{Salo:88,Bridges:84}.  But surprisingly, such model has
not yet been tested numerically for strongly vibrated granular media. 

\begin{figure}[htb]
\begin{center}
\includegraphics[width=.67\textwidth]{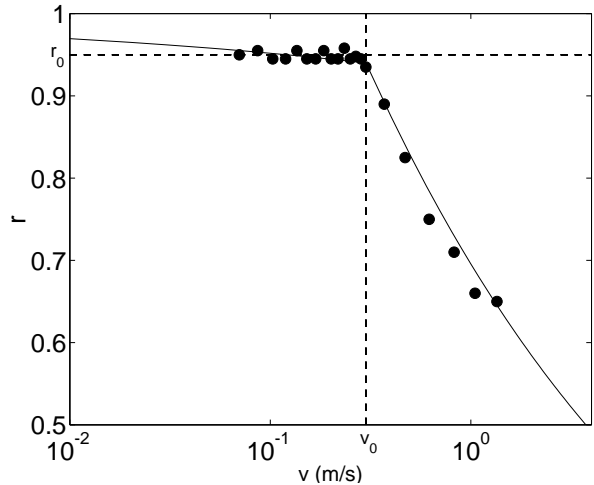}
\end{center}
\caption[]{The restitution coefficient $r$ as
 a function
of impact velocity $v$, as given in Eq.~(\ref{McNFal:eq:rvimp}) (solid line).  The dashed lines
 show $v_0=0.3$ m/s and $r_0=0.95$. Experimental points ($\bullet$) for
steel spheres were extracted from Fig.1 of Ref. \cite{Lifshitz:64}}
\label{McNFal:fig:rvimp}
\end{figure}
In this paper,
we use a velocity dependent restitution coefficient  $r(v)$ and join the
two regimes of dissipation (viscoelastic and plastic) together as simply as
possible, assuming that
\begin{equation}
r(v) = \left\{  \begin{matrix}   
  1-(1-r_0) \left(\frac{v}{v_0} \right)^{1/5}, &  v \le v_0, \cr
     r_0
\left( \frac{v}{v_0} \right)^{-1/4}, & v \ge v_0 ,\end{matrix}\right.
\label{McNFal:eq:rvimp}
\end{equation}
where $v_0 = 0.3$ m/s is chosen, throughout the paper, to be the yielding
velocity for stainless steel particles \cite{Johnson:85,Lifshitz:64} for
which $r_0$ is close to 0.95 \cite{Lifshitz:64}. Note that $v_0 \sim 1/\sqrt{\rho}$ where $\rho$ is the density of the sphere \cite{Johnson:85}. We display in
Fig.~\ref{McNFal:fig:rvimp} the velocity dependent restitution coefficient of
Eq.~(\ref{McNFal:eq:rvimp}), with $r_0=0.95$ and $v_0 = 0.3$ m/s, that agrees well
with experimental results on steel spheres from Ref.\ \cite{Lifshitz:64}. As also already noted by Ref.\ \cite{Johnson:85}, the impact velocity to cause yield in metal surfaces is indeed relatively small. 
For metal, it mainly comes from the low yield stress value ($Y \sim 10^9$ N/m$^2$) with repect to the elastic the Young modulus ($E \sim 10^{11}$ N/m$^2$). Most impacts between metallic bodies thus involve some plastic deformation. 

\subsection{The other simulation parameters}

The numerical simulation consists of an ensemble of identical hard disks
of mass $m \approx 3 \times 10^{-5}$ kg
excited vertically by a piston in a two-dimensional box. Simulations are
done both in the presence and absence of uniform gravity $g$. Collisions
are assumed instantaneous and thus only binary collisions occur. For
simplicity, we neglect the rotational degree of freedom.  
Collisions with the wall are treated in the same way as collisions between
particles, except the wall has infinite mass. 

Motivated by recent 3D experiments on stainless steel spheres, 2 mm in
diameter, fluidized by a vibrating piston \cite{Falcon:01}, we choose the
simulation parameters to match the experimental ones: in the simulations, the vibrated piston
at the bottom of the box has amplitude $A=25$ mm and frequencies
$5\;\text{Hz} \le f \le 50\;\text{Hz}$.  The piston is nearly sinusoidally vibrated with
a waveform made by joining two parabolas together, leading to a maximum piston velocity given by $V=4Af$.
The particles are disks 
$d = 2\;\text{mm}$
in diameter with stainless steel collision properties through $v_0$ and
$r_0$ (see Fig.~\ref{McNFal:fig:rvimp}). 
The box
has width $L=20$ cm and horizontal periodic boundary conditions. 
Since our simulations are two
dimensional,  we consider the simulation geometrically equivalent to the
experiment when their number of layers of particles, $n=Nd/L$, are equal. 
Hence in the simulation, a layer of particles, $n=1$, corresponds to $100$
particles.  We checked that $n$ is an appropriate way to measure
the number of particles by also running simulations at $L=10$ cm
and $L=40$ cm.  None of this paper's results depend significantly
on $L$, excpet in Sec.~\ref{McNFal:sec:gcluster}, where this point 
is discussed.
As in the experiments, the height $h$ of the box depends on the
number of particles in order to have a constant difference $h-h_0=15$ mm,
where $h_0$ is the height of the bed of particles at rest. Heights are
defined from the piston at its highest position.

\section{Comparison of simulation and experiment}
\label{McNFal:sec:experiment}

\subsection{The importance of the variable coefficent of restitution}

\begin{figure}[htb]
\begin{center}
\begin{tabular}{cc}
{\bf (a) experiments} & {\bf (b) $r=r(v)$}\\
\includegraphics[width=.49\textwidth]{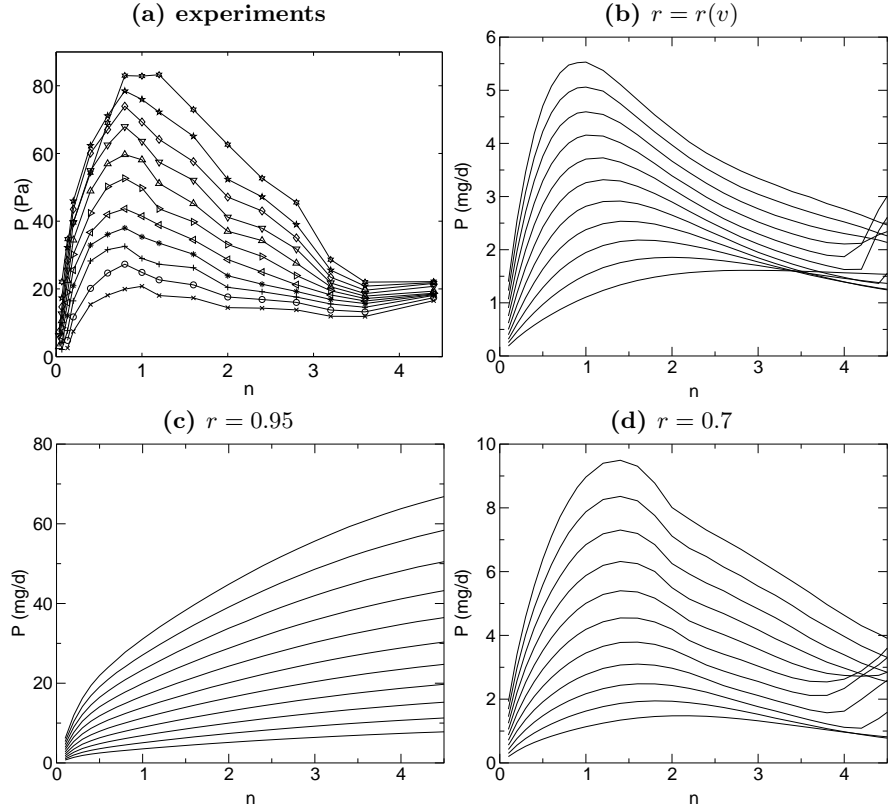}&
\includegraphics[width=.49\textwidth]{fig02b.eps}\\
{\bf (c) $r=0.95$} & {\bf (d) $r=0.7$}\\
\includegraphics[width=.49\textwidth]{fig02c.eps}&
\includegraphics[width=.49\textwidth]{fig02d.eps}\\
\end{tabular}
\end{center}
\caption[]{Time averaged pressure $P$ on the top of the cell 
as a function of particle layer, $n$, for various vibration frequencies,
$f$ : {\bf (a)} Experimental results from \cite{Falcon:01} for stainless
steel beads 2 mm in diameter, with $A=25$ mm, $10\le f \le 20 Hz$ with a 1
Hz step (from lower to upper)  and $h-h_0 = 5$ mm. {\bf (b)} Numerical
simulation where the coefficient of restitution is given by
Eq.~(\ref{McNFal:eq:rvimp}), {\bf (c)} Numerical simulation with a coefficient of
restitution is $0.95$, independent of impact velocity,
{\bf (d)} Numerical simulation with a coefficient of
restitution is $0.7$, independent of impact velocity. The simulations
(b) -- (d) are 2D with gravity, done for 2 mm disks,
with $A=25$ mm, $10\le f \le 30$ Hz with a 2 Hz step
(from lower to upper) and $h-h_0 = 15$ mm.  In the simulations, the
two-dimensional pressure is given in units of 
$mg/d \approx 0.15\;\mathrm{Pa}\cdot\mathrm{m}$.}
\label{McNFal:fig:freqscan}
\end{figure}

We examine first the dependence of the pressure on
the number of particle layers for maximum velocity of the piston
$1 \lesssim V\lesssim 5$ m/s ($V=4Af$). The time averaged pressure at
the upper wall is displayed in Fig.~\ref{McNFal:fig:freqscan} as a function of $n$
for various $f$: from the experiments of Falcon et al. \cite{Falcon:01} (see
Fig.~\ref{McNFal:fig:freqscan}a), from our simulations with velocity dependent
restitution coefficient $r=r(v)$ proposed in Eq.~(\ref{McNFal:eq:rvimp}) (see
Fig.~\ref{McNFal:fig:freqscan}b), and with constant restitution coefficient
$r=0.95$, often used to describe steel particles (see
Fig.~\ref{McNFal:fig:freqscan}c), and finally with an unrealistic constant restitution coefficient
$r=0.7$ (see Fig.~\ref{McNFal:fig:freqscan}d).  
Simulations with $r=r(v)$ give results in
agreement with the experiments: At constant external driving, the pressure
in both Figs.~\ref{McNFal:fig:freqscan}a and \ref{McNFal:fig:freqscan}b passes through a
maximum for a critical value of $n$ roughly corresponding to one particle
layer.  For $n<1$, most particles are in vertical ballistic motion between
the piston and the lid. Thus, the mean pressure increases roughly
proportionally to $n$. When $n$ is increased such that $n>1$, interparticle
collisions become more frequent. The energy dissipation is increased and
thus the pressure decreases.  As we show below in Sec.~\ref{McNFal:sec:g0clusters},
this maximum pressure is not due to
gravity because it also appears in simulations with $g=0$ and $r=r(v)$.
Turning our attention to Fig.~\ref{McNFal:fig:freqscan}c, we see
that setting $r=0.95$ independently of impact velocity gives pressure
qualitatively different from experiments. The difference between
Figs.~\ref{McNFal:fig:freqscan}b and \ref{McNFal:fig:freqscan}c can be understood by
considering a high velocity collision (e.g. $v=1$ m/s). In
Fig.~\ref{McNFal:fig:freqscan}b, this collision has a restitution coefficient of
$r=r(1$ m/s$) \approx 0.7$ (see Fig.~\ref{McNFal:fig:rvimp}), whereas in
Fig.~\ref{McNFal:fig:freqscan}c, $r$ is fixed at 0.95 for all collisions. This
means that for equal collision frequencies, dissipation is much stronger
for $r=r(v)$ than for $r=0.95$, because the high velocity collisions
dominate the dissipation. Stronger dissipation leads to lower granular
temperatures and thus to lower pressures. 

We can check this interpretation
by changing the constant restitution coefficient to $r=0.7$ and then
comparing it to $r=r(v)$. In these two cases, the high velocity collisions
will have roughly the same restitution coefficient. We indeed observed a
pressure that decreases for large $n$ for constant $r=0.7$
(see Fig.~\ref{McNFal:fig:freqscan}d). 
Therefore, surprisingly, constant $r=0.7$ reproduces more precisely the experimental
pressure measurements than constant $r=0.95$, even though $r=0.95$ or
$r=0.9$ is often given as the restitution coefficient of steel.
Consulting Fig.~\ref{McNFal:fig:rvimp}, one can see that $r(v)=0.7$ is 
well inside the plastic
regime.  Therefore, our results suggest that plasticity is much more
important than visco-elasticity in vibrated granular media.
Although constant $r=0.7$ gives good agreement with the experiments
in Fig.~\ref{McNFal:fig:freqscan}, this will not be true in all situations.
If the inter-particle collisions were much gentler, the effective
coefficient of restitution would rise, and another choice of constant
$r$ would be needed.  Furthermore, the pressure is just one property
of the system.  If we look at other property, we see that $r=0.7$ does
not give good results.  
\begin{figure}
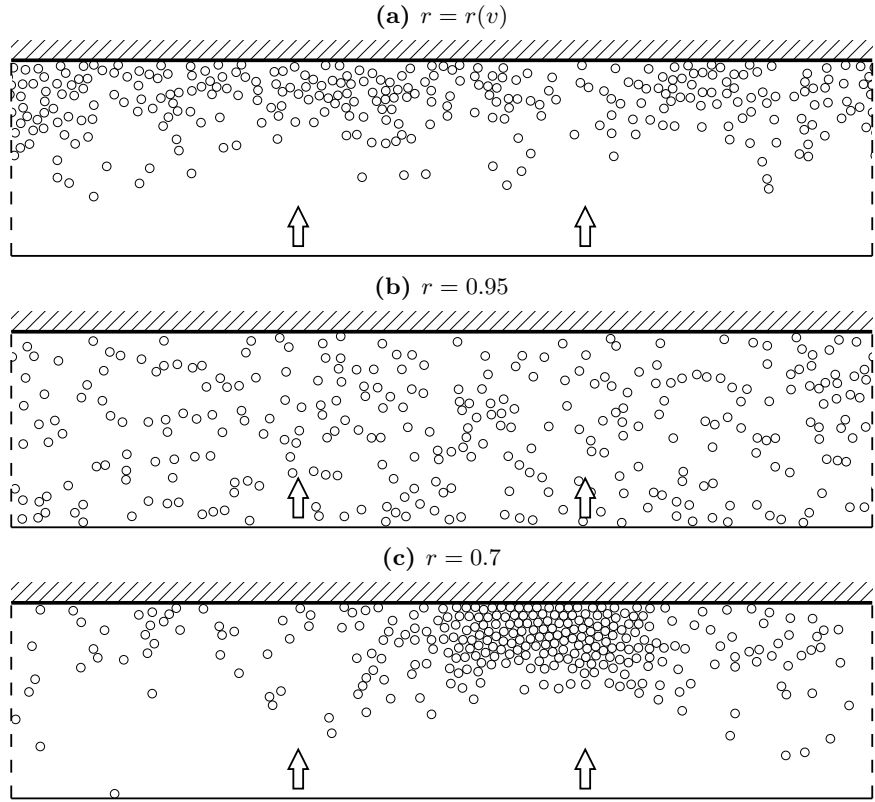

\begin{center}
\begin{tabular}{c}
{\bf (a) $r=r(v)$} \\
\includegraphics[width=0.99\textwidth]{fig03a.eps}\\
{\bf (b) $r=0.95$} \\
\includegraphics[width=0.99\textwidth]{fig03b.eps}\\
{\bf (c) $r=0.7$} \\
\includegraphics[width=0.99\textwidth]{fig03c.eps}\\
\end{tabular}
\end{center}
\caption[]{Snapshots from the simulations with $n=3$, gravity $g\ne0$, driving frequency $f=30\;\text{Hz}$ and $h-h_0=15\;\text{mm}$.
The upper wall is stationary, and the lower wall is the piston,
which is at its lowest position in all three snapshots.  The horizontal
boundaries are periodic (indicated by dashed lines).  Gravity points
downwards.  {\bf (a)} $r=r(v)$, as given in Eq.~(\ref{McNFal:eq:rvimp}),
{\bf (b)} constant $r=0.95$, and {\bf (c)} constant $r=0.7$.  
In (c) we see a tight cluster which was not observed in the experiments.}
\label{McNFal:fig:snapshots}
\end{figure}
An example is shown in Fig.~\ref{McNFal:fig:snapshots},
where we show three snapshots from three different simulations with
$n=3$ in absence of gravity.  When $r=r(v)$, the particles are concentrated in the upper half
of the chamber, but they are evenly spread in the horizontal direction (see Fig.~\ref{McNFal:fig:snapshots}a). 
The system is hotter and less dense near the vibrating wall, and colder and denser by the opposite wall. Such a loose cluster has been already observed in microgravity during parabolic flight experiments \cite{Falcon:unpub}, where grains are agitated with a piston.
In contrast, when $r=0.95$, the particles are uniformly distributed
throughout the domain (see Fig.~\ref{McNFal:fig:snapshots}b).  This occurs because the particle move much faster
than the piston, and thus can fill the space left by the descending piston.
Finally, when $r=0.7$, the majority of the particles are confined to
a tight cluster, pressed against the upper wall, coexisting with low density region (see Fig.~\ref{McNFal:fig:snapshots}c).  
This instability has been already
been reported numerically \cite{Argentina:02}, although for much different
parameters (constant restitution coefficient $r=0.96$, thermal walls, 
no gravity, and large $n$). However, nothing like this was
ever seen experimentally. Therefore, if one
is seeking information about particle positions, $r=0.7$ gives incorrect
results even though it gives acceptable results for the pressure.
We conclude,
therefore, that the only way to successfully describe all the properties
in all situations is to use a velocity dependent restitution
coefficient model.  

\subsection{The importance of the particle number}

\begin{figure*}
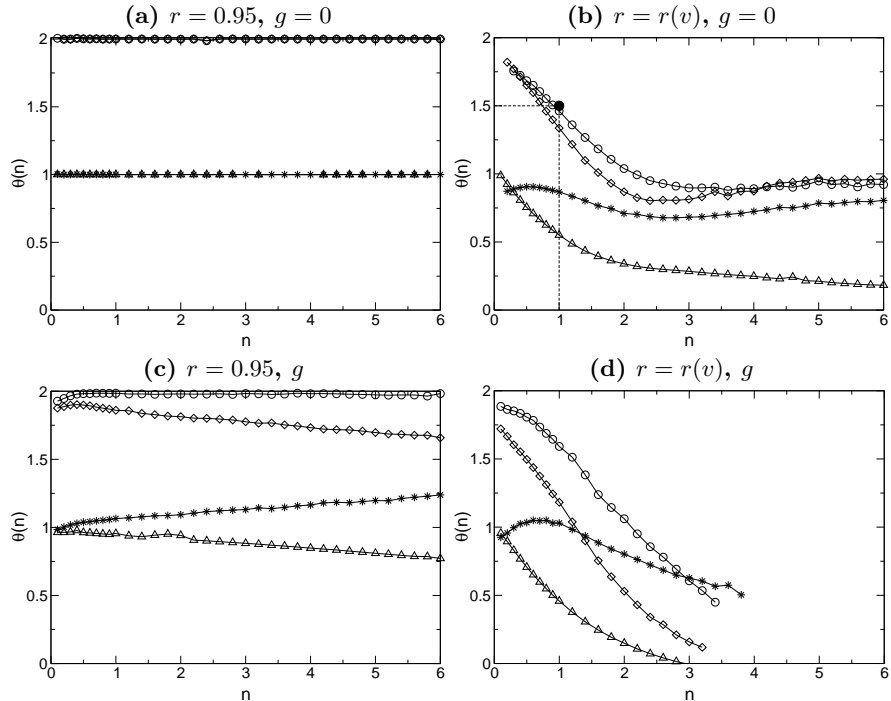

\begin{center}
\begin{tabular}{cc}
{\bf (a) $r=0.95$, $g=0$} & {\bf (b) $r=r(v)$, $g=0$}\\
\includegraphics[width=.49\textwidth]{fig04a.eps} &
\includegraphics[width=.49\textwidth]{fig04b.eps} \\
{\bf (c) $r=0.95$, $g$ } & {\bf (d) $r=r(v)$, $g$}\\
\includegraphics[width=.49\textwidth]{fig04c.eps} &
\includegraphics[width=.49\textwidth]{fig04d.eps} \\
\end{tabular}
\end{center}
\caption{ The exponents  $\theta$ as a
function of $n$ which give the scaling of the granular temperature $T$
($\Diamond$), collision frequency $C_{\text{upper}}$ ($\ast$), mean impulsions $\Delta
I$ ($\bigtriangleup$) and pressure $P$ ($\circ$). All these quantities are
proportional to $V^{\theta(n)}$.  Without gravity: {\bf (a)} for $r=0.95$
and {\bf (b)} for $r=r(v)$. With gravity: {\bf (c)} for $r=0.95$ and {\bf
(d)} for $r=r(v)$.  The exponents are obtained by fixing $n$ and
performing eleven simulations, varying $f$ from $10$ Hz to $30$ Hz.
Then $\log X$ (where $X$ is the quantity being considered) is plotted
against $\log V$.  The resulting curve is always nearly a straight
line (except for $n>3$ in (d) -- see text),  and the exponent is 
calculated from a least squares fit.   
The pressure scaling point ($\bullet$) on (b)
is from experiment \cite{Falcon:99} performed in low-gravity. See Fig.~\ref{McNFal:fig:snapshots}a (resp. Fig.~\ref{McNFal:fig:snapshots}b) for typical snapshots corresponding to $n=3$, $g\ne0$ and $r=r(v)$ (resp. $r=0.95$)}
\label{McNFal:fig:thetap}
\end{figure*}

Observations suggest that the  the pressure (or granular temperature) of strongly vibrated granular medium is controlled
by the piston vibration velocity $V$. It is important to 
point out that $V$ is not
the only way to characterize the vibration.  One could also use the
maximum piston acceleration $\Gamma$.  When $\Gamma$ is close to $g$,
it controls the behavior of the system, i.e., adjusting $A$ and $f$
while keeping $\Gamma$ constant does not change the system's behavior
much.  But in the simulations presented here, $\Gamma\gg g$, and the
system's behavior is controlled by $V$.  This can be checked by
multiplying the frequency by $10$ while dividing $A$ by $10$,
thus keeping $V$ the same (while $\Gamma$ increases by an order
of magnitude).  Doing so changes the pressure only by about $20\%$.
It is reasonable, therefore, to look for scaling relations in $V$.

Many authors have postulated that the pressure on the upper wall $P$ (or
granular temperature $T$) is related to the piston velocity $V$
through $P,T \propto V^\theta$.  However, it is not clear what the
correct ``scaling exponent'' $\theta$ should be.  This question
 has been
addressed several times in the past, without a clear resolution of the
question \cite{McNamara:98,Huntley:98,Kumaran:98,Lee:95}. For example, kinetic
theory \cite{Warr:95,Kumaran:98} and hydrodynamic models \cite{Lee:95} predict
$T \propto V^{2}$ whereas numerical simulations \cite{Luding:94a,Luding:94b}
or experiments \cite{Warr:95,Yang:00,Luding:94a} give $T \propto V^{\theta}$,
with $1 \leq \theta \leq 2$. These studies were done at single values of
$n$.  In this section, we show that it is very important to
explicitly consider the dependence of the scaling exponents on $n$. 
We also consider the effect of gravity and a variable coefficient
of restitution.
Doing so enables us to explain and unify all previous works.

At the upper wall, we measured numerically the collision frequency, 
$C_{\text{upper}}$ and the mean impulsion per
collision, $\Delta I$ for various
frequencies of the vibrating wall and numbers of particles in the box, with
$r=r(v)$ or with $r=0.95$, in the presence or absence of gravity.
The time averaged pressure on the upper wall
can be calculated from these quantities
using  
\begin{equation}
P=C_{\text{upper}} \Delta I / L \ {\rm .}
\label{McNFal:eq:defP}
\end{equation}
(By conservation of momentum, the time averaged pressure on the
lower wall is just $P$ plus the weight of the particles $Nmg$.)
The total kinetic energy of the system is also measured to have access 
to the granular temperature, $T$. $C_{\text{upper}}$, $\Delta I$, $P$,
and $T$ are all found
to fit with power laws in $V^{\theta}$ for our range of piston velocities.
Fig.\ \ref{McNFal:fig:thetap} shows $\theta$ exponents of $C_{\text{upper}}$, 
$\Delta I$, $P$,
and  $T$ as a function of $n$. When $g=0$ and $r$ is constant (see Fig.\
\ref{McNFal:fig:thetap}a), we have $P\sim V^2$, $\Delta I \sim V$ 
and $C_{\text{upper}} \sim V$
for all $n$. We call these relations the classical kinetic theory scaling.
This scaling can be established by simple dimensional analysis when the
vibration velocity $V$ provides the only time scale in the system. This is
the case for $g=0$ and $r$ independent of velocity. However, in the
experiments, two additional time scales are provided, one by gravity and
another by velocity dependent restitution coefficient. Numerical simulations
can separate the effects of these two new time scales on the scaling
exponents $\theta$. This is done in Fig.\ \ref{McNFal:fig:thetap}b [where $g=0$
but $r=r(v)$] and Fig.\ \ref{McNFal:fig:thetap}c [where $r$ is constant but
$g\ne0$]. In both figures, all the exponents become functions of $n$.
However, the time scale linked to $r=r(v)$ leads to much more dramatic
departure from the classical scaling. After considering the two time scale
separated, let us consider the case corresponding to most experiments,
where both gravity and $r=r(v)$ are present (see Fig.\ \ref{McNFal:fig:thetap}d).
The similarity between this figure and Fig.\ \ref{McNFal:fig:thetap}b confirms
that the velocity dependent restitution coefficient has a more important
effect than the gravity. Furthermore, only the variation of the restitution
coefficient on the particle velocity explains the experiment performed in
low-gravity \cite{Falcon:99}. This experiment gives a $V^{3/2}$ pressure
scaling ($\bullet$-mark on Fig.\ \ref{McNFal:fig:thetap}b) for $n=1$ and a
motionless clustered state for $n>2$. Only the simulation with $r=r(v)$ can
reproduce these results (see Fig.\ \ref{McNFal:fig:thetap}b) whereas constant $r$
simulations leads to the classical scaling ($P \propto V^{2}$, see Fig.\
\ref{McNFal:fig:thetap}a) and only a gaseous state for all $n$ shown in the figure.

\begin{figure*}
\begin{center}
\includegraphics[width=.67\textwidth]{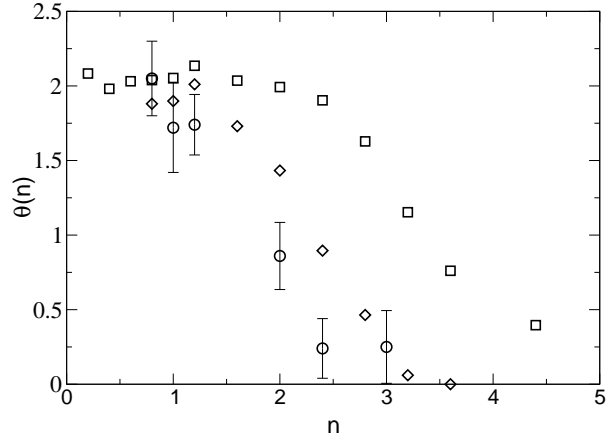}
\end{center}
\caption[]{Experimental data performed under gravity from Ref. 
 \cite{Falcon:01}: The
exponents $\theta(n)$ of time averaged pressure ($\square$) (see
Fig.~\ref{McNFal:fig:freqscan}a), and kinetic energy extracted from
density profile ($\circ$) or volume expansion ($\lozenge$) measurements. 
These data should be compared with the simulations of 
Fig.~\ref{McNFal:fig:thetap}d.}
\label{McNFal:fig:eric}
\end{figure*}

As shown in Fig\ \ref{McNFal:fig:thetap}, 
it is thus very important to explicitly consider the dependence of $\theta$
on $n$.  In all cases, except the unrealistic case of Fig.\
\ref{McNFal:fig:thetap}a, $\theta$ depends on $n$.  To our
knowledge, the only experiment  \cite{Falcon:01} to systematically
investigate this effect shows that  $T \propto V^{\theta(n)}$, with $\theta$
continuously varying from $\theta = 2$ when $n \rightarrow 0$, as expected
from kinetic theory, to $\theta \simeq 0$ for large $n$ due to the
clustering instability. These experiments \cite{Falcon:01} performed under
gravity (shown in Fig.\ \ref{McNFal:fig:eric}) are well reproduced
by the simulations of Fig.\ \ref{McNFal:fig:thetap}d. In both cases, the observed
pressure and granular temperature scaling exponents strongly decrease with
$n$. 

We finish this section by noting two curious facts about
Fig.~\ref{McNFal:fig:thetap}.  First of all, in Fig.~\ref{McNFal:fig:thetap}b
[$g=0$ and $r=r(v)$], $\theta\approx 1$ for the pressure and temperature
when $n>2$.  This is the sign of new robust scaling regime where $P,T \propto V$,
which is the subject of Sec.~\ref{McNFal:sec:g0clusters}.
Secondly, in Fig.~\ref{McNFal:fig:thetap}d [$g \neq 0$ and $r=r(v)$], the scaling exponents are not
shown for $n\ge3$, because the dependence of $P$, $T$ $C_{\text upper}$
and $\Delta I$ on $V$ is no longer a simple power law.  (More
precisely, we do not plot a point on
Fig.~\ref{McNFal:fig:thetap} when
$\left| \log X_{\text{observed}}-\log X_{\text{fitted}}\right| \ge 0.25$
for any of the eleven simulations used to calculate the exponent --
see caption.)
The power law breaks down because there is a resonance
between the time of flight of the cluster under gravity and the 
vibration period. This is the subject of Sec.~\ref{McNFal:sec:gcluster}.

\section{Effects of clustering ($n>2$ or 3)}
\label{McNFal:sec:clusters}

\subsection{Clustering without gravity}
\label{McNFal:sec:g0clusters}

\begin{figure*}
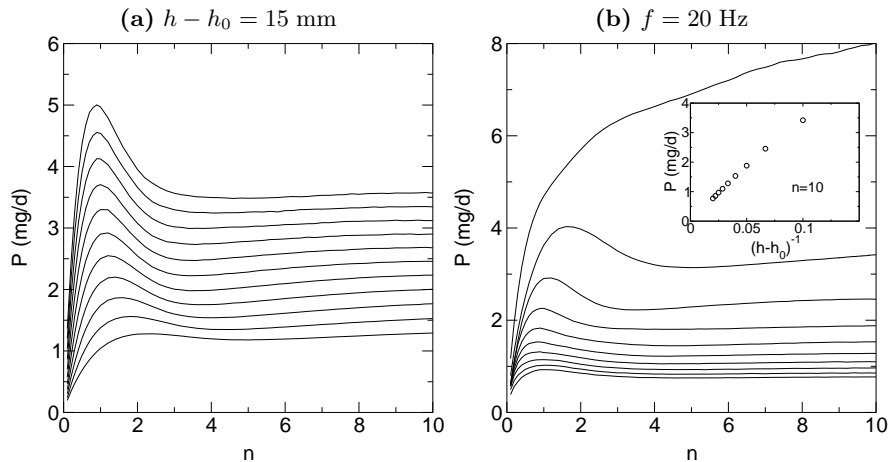

\begin{center}
\begin{tabular}{cc}
{\bf (a)} $h-h_0=15\;\text{mm}$ & {\bf (b)} $f=20\;\text{Hz}$\\
\includegraphics[width=.49\textwidth]{fig06a.eps} &
\includegraphics[width=.49\textwidth]{fig06b.eps}
\end{tabular}
\end{center}
\caption{Time averaged pressure $P$ on the top of the cell 
as a function of particle layer, $n$:
{\bf (a)} for various vibration frequencies: $f=10\;\text{Hz}$ (lowest curve)
to $f=30\;\text{Hz}$ (uppermost curve) with steps of 2 Hz; {\bf (b)} for various heights:
$h-h_0=5\;\text{mm}$ (uppermost curve) to $h-h_0=50\;\text{mm}$
(lowest curve) with steps of 5 mm.  Inset: $P$ vs.~$(h-h_0)^{-1}$ for $n=10$ and
$10\;\mathrm{cm} \le h-h_0 \le 50\;\mathrm{cm}$.
Simulations were done with $g=0$ and $r=r(v)$.}
\label{McNFal:fig:g0rvimp}
\end{figure*}

In this section, we examine the situation concerning realistic particles 
[$r=r(v)$] in
microgravity ($g=0$).  At high enough density, a loose cluster is formed
that resembles Fig.~\ref{McNFal:fig:snapshots}a (even though $g\ne0$ in
Fig.~\ref{McNFal:fig:snapshots}a), and we noted that $P \propto V$ for
$n>2$ (see Fig.~\ref{McNFal:fig:thetap}b).  We would like to know if there is
some simple physical explanation for this aparently robust scaling
relation.  To investigate this question, we present in
Fig.~\ref{McNFal:fig:g0rvimp}
the dependence of $P$ on $V$, $n$ and $h-h_0$.
Fig.~\ref{McNFal:fig:g0rvimp}a shows the same quantities as Fig.~\ref{McNFal:fig:freqscan},
except that the scale of the $x$-axis has changed: instead of $0 < n < 4.5$,
we show $0 < n < 10$.  Note that Fig.~\ref{McNFal:fig:g0rvimp}a concerns the same
situation as in Fig.~\ref{McNFal:fig:freqscan}b, except here the gravity has
been ``turned off''.  Thus it is not surprising that for $n<3$, the
two figures are almost the same: after a rapid increase of the pressure
for small $n$, there is a maximum pressure near $n\approx1$.
But for $n>3$, the figures are different: in Fig.~\ref{McNFal:fig:freqscan}b,
the pressure decreases as particles are added, but in
Fig.~\ref{McNFal:fig:g0rvimp}a the pressure is nearly independent of $n$.
Therefore,
when $g=0$, the pressure is not changed by adding more particles,
as long as there are already enough particles present.  Considering
now the dependence of $P$ on $V$, we note that the evenly spaced
curves in Fig.~\ref{McNFal:fig:g0rvimp}a suggest that $P \propto V$.
We have confirmed this proportionality by calculating the pressure
scaling exponent and
checking that it remains close to $1$.  (The exponent for $n<6$
has already been shown in Fig.~\ref{McNFal:fig:thetap}b.)
In Fig.~\ref{McNFal:fig:g0rvimp}b,
we examine how changing $h-h_0$ affects the pressure.  We see that
the pressure
is independent of $n$ for $n>3$ and $h-h_0\ge10$, 
with the pressure decreasing
as the height increases.  Examining the dependence of the pressure
on $h-h_0$ shows that $P \propto (h-h_0)^{-1}$.  (For example, see
the inset in Fig.~\ref{McNFal:fig:g0rvimp}b.)
Thus, the pressure
obeys the simple non-extensive relation
\begin{equation}
P \propto \frac{N^0 V^1}{h-h_0}
\label{McNFal:eq:Pscale}
\end{equation}
Recall that this relation concerns high enough number of
realistic particles in microgravity.
Therefore, this pressure scaling of Eq.~(\ref{McNFal:eq:Pscale}) may be observable in
microgravity experiments.  One complication is that in microgravity
experiments, one does not usually agitate the grains with a piston (except very recently \cite{Falcon:unpub}) --
it is much simpler to shake a box of fixed size \cite{Falcon:99}.

\begin{figure*}
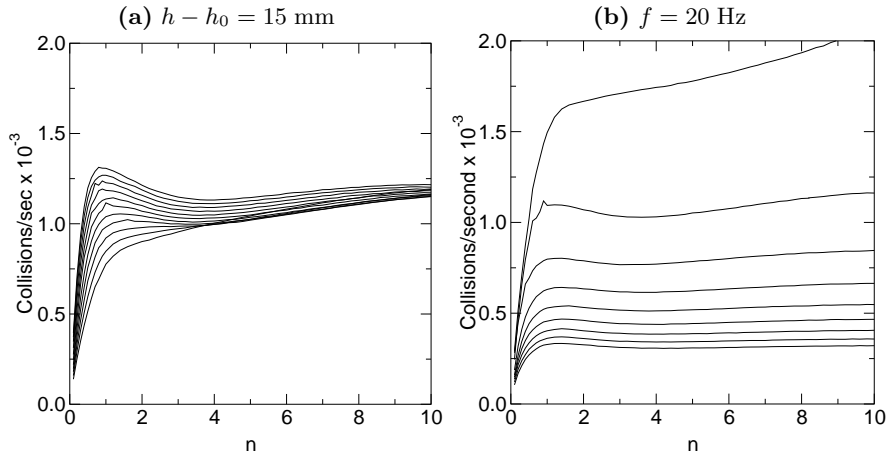

\begin{center}
\begin{tabular}{cc}
{\bf (a)} $h-h_0=15\;\text{mm}$ & {\bf (b)} $f=20\;\text{Hz}$\\
\includegraphics[width=.49\textwidth]{fig07a.eps} &
\includegraphics[width=.49\textwidth]{fig07b.eps}
\end{tabular}
\end{center}
\caption{The particle-piston collision rate $C_{\text{lower}}$ 
as a function of particle layer, $n$:
{\bf (a)} for various vibration frequencies: $f=10\;\text{Hz}$ (lowest curve)
to $f=30\;\text{Hz}$ (uppermost curve) with steps of $2$ Hz; 
{\bf (b)} for various heights:
$h-h_0=10\;\text{mm}$ (uppermost curve) to $h-h_0=50\;\text{mm}$
(lowest curve) with steps of $5$ mm. 
Simulations are done with $g=0$ and $r=r(v)$.}
\label{McNFal:fig:Cdn}
\end{figure*}

In Fig.~\ref{McNFal:fig:snapshots}a, we observe that the majority of particles 
remain in a loose cluster pushed
against the stationary wall, opposite the piston.  Only those particles
that ``evaporate'' from the cluster are struck by the piston.  The
evaporation rate can be estimated from the rate of collisions between 
the piston and the particles $C_{\text{lower}}$.  This collision
rate has a very curious behavior, as shown in Fig.~\ref{McNFal:fig:Cdn}.
As can be seen from Fig.~\ref{McNFal:fig:Cdn}a, 
$C_{\text{lower}}$ is roughly independent of $n$ and $V$ when $n>3$.
This behavior hold for other values of $h-h_0$,
as can be seen from Fig.~\ref{McNFal:fig:Cdn}b.  Thus, at high enough density, the collision frequency on the vibrating wall is found to be
\begin{equation}
C_{\text{lower}} \propto \frac{N^0 V^0}{h-h_0}.
\label{McNFal:eq:Cscale}
\end{equation}

We now present a simple model and a physical interpretation of the 
unexpected scaling of Eq.~(\ref{McNFal:eq:Pscale}) for such a loose cluster.  
We suppose that the cluster against the
stationary wall has a granular temperature independent of both the
number of particles $N$ and the vibration velocity $V$.  
The granular temperature is independent of $N$ and $V$ because
the impact velocity dependent restitution coefficient acts as a
kind of ``granular thermostat''.  As one can see from
Eq.~(\ref{McNFal:eq:rvimp}) or Fig.~\ref{McNFal:fig:rvimp},
when the typical particle
velocity $v$ is greater than $v_0$, $r(v)$ is small, and energy 
is dissipated rapidly.  On the other hand, when $v<v_0$, $r(v)$ is nearly
unity and energy dissipation is slow.  Thus, in a weakly forced cluster,
like the one found at the stationary wall, $T\sim v^2$ will be near
$v_0^2$, independent of the strength of the forcing and the
size of the cluster.  

Since the granular temperature of the cluster is independent of
$N$ and $V$, so is the
flux $F_{\text{evap}}$ of grains ``evaporating'' from
the cluster and moving towards the piston.  If all the grains that
evaporate from the cluster reach the piston, then we have
$C_{\text{lower}} = F_{\text{evap}} \propto N^0V^0$.  In their
collision with the piston, the grains acquire an upwards velocity
proportional to $V$.  Thus the flux of momentum entering the system
at the piston is $V C_{\text{lower}} \propto  N^0V^1$.  Since momentum
is conserved, the flux of momentum leaving the system through the
stationary wall must have the same value.  But the time averaged
pressure $P$ is just the time averaged momentum flux divided by
the area of the upper wall.  Thus our model predicts
$P \propto  N^0V^1$, in agreement with Eq.~(\ref{McNFal:eq:Pscale}).

This theory also explains why $P$ and $C_{\text{lower}}$ are
inversely proportional to $h-h_0$.  When a particle evaporates from
the cluster, it must travel a certain distance before it encouters
the piston.  This distance increases with $h-h_0$ and thus the
particle's travel time also increases.  During its voyage, the
evaporated particle could be struck by another particle that
has just encountered the piston.  If this happens, both particles
are scattered back into the cluster.  Thus the evaporated particle
never reaches the piston.  If we assume that the probability of
an evaporated particle being scattered back into the cluster is
independent of time, the number of particles reaching the piston
is inversely proportional to $h-h_0$.

\begin{figure}
\begin{center}
\includegraphics[width=.67\textwidth]{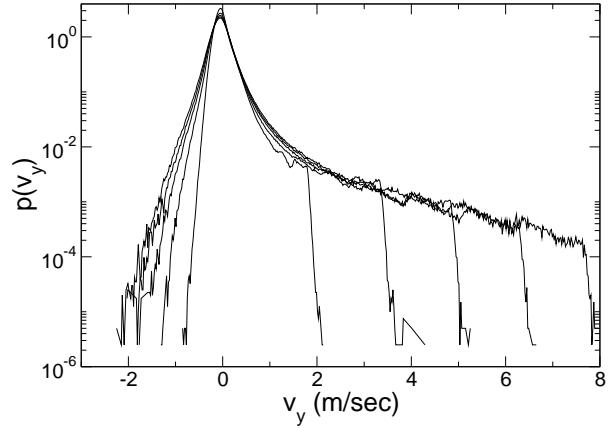}
\end{center}
\caption[]{The probability density function of the vertical velocities
at high density, for $n=5$, $r=r(v)$, and $g=0$.  The driving frequencies are
$f=10\;\text{Hz}$ (narrowest distribution) to $f=50\;\text{Hz}$
(broadest distribution) with steps of 10 Hz.}
\label{McNFal:fig:vhist}
\end{figure}

This explanation can be checked by looking at the probability density of
vertical ($y$) velocities parallel to the direction of vibration.  In Fig.~\ref{McNFal:fig:vhist}, 
we show these density functions.  Note the assymmetry of the distibution's
positive and negative wings.  The edge of the positive wing increases
linearly with $V$ whereas the negative wing extends much more slowly.
The reason for this is that the positive wing is populated by particles
which have just been struck by the piston.  Their velocity is thus 
proportional to $V$.  On the other hand, the negative wing is due
to fluctuations in the cluster.  The negative wing thus expands slowly
because of the ``thermostat'' effect described above.

We conclude by emphasizing that this effect can only be seen with the
velocity dependent restitution coefficient.  If one uses a constant
restitution coefficient, one obtains $P \propto V^2$, no matter the
value of $r$.  

\subsection{Clustering with gravity}
\label{McNFal:sec:gcluster}

In this section, we consider the situation that applies to most
of the experiments, i.e.~realistic particles [$r=r(v)$]
under gravity.  We will show that the relatively simple situation
described in the previous section is complicated by the presence
of gravity.  Under gravity, the loose cluster will not remain near the
upper wall (unless $f$ is large),
but will fall downwards towards the piston.  At certain
values of $f$, $h-h_0$ and large enough $n$, a resonance occurs
between the driving frequency and the time of flight of the particles,
leading to a cluster that bounces back and forth between the piston
and the wall.  As $f$ increases, however, we recover the situation
found in the previous section.

\begin{figure*}
\begin{center}
\includegraphics[width=.99\textwidth]{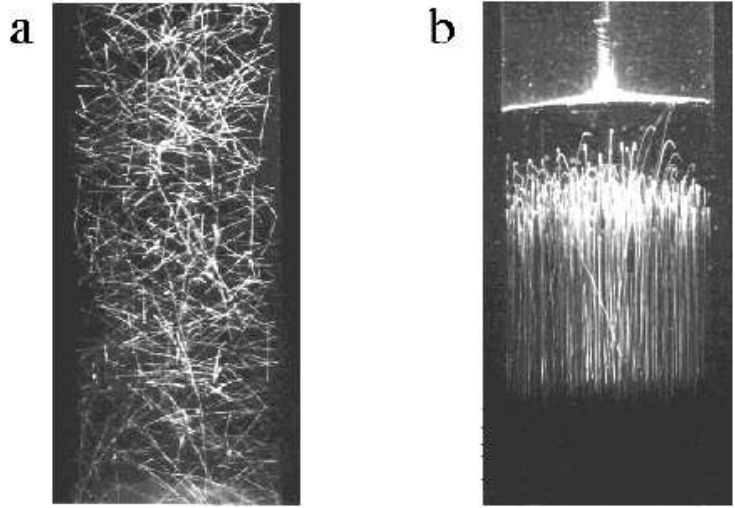}
\end{center}
\caption[]{Photographs of particle trajectories from the experiments
of \cite{Falcon:01}, showing the cluster at high densities.  
In both pictures $f=20\;\text{Hz}$ and $A=40\;\text{mm}$.  The
number of layers of particles is {\bf(a)} $n=1$ {\bf (b)} $n=4$. The driving piston is at the bottom (not visible), the inner diameter of the tube being 52 mm.}
\label{McNFal:fig:photos}
\end{figure*}

\begin{figure*}
\begin{center}
\begin{tabular}{ccc}
{\bf (a) $n=1$} &$\quad$& {\bf (b) $n=5$}\\
\includegraphics[width=.4\textwidth]{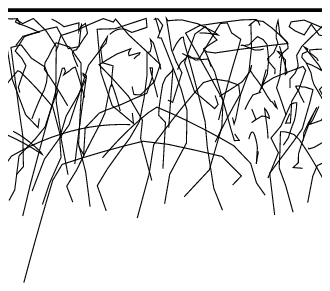} &&
\includegraphics[width=.4\textwidth]{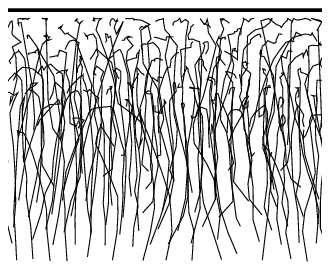}\\
\end{tabular}
\end{center}
\caption{Particle trajectories from the simulations with 
  $f=20\;\text{Hz}$, $A=25$ mm, $r=r(v)$, and $h-h_0=15\;\text{mm}$.
The number of layers of particles is {\bf (a)} $n=1$ and 
{\bf (b)} $n=5$.  The containers have different sizes, because
the size of the container is increased as more particles are added,
as was done in the experiments.}
\label{McNFal:fig:simus}
\end{figure*}

In Fig.~\ref{McNFal:fig:photos}, we show two pictures from the experiments of 
Ref.~\cite{Falcon:01}.  At low densities, the particles
move like atoms in a gas (Fig.~\ref{McNFal:fig:photos}a).   At higher
densities, a cluster forms that bounces up and down on the
piston (Fig.~\ref{McNFal:fig:photos}b).  This cluster was observed only for
certain frequencies, densities and container heights.  It was not
possible to establish a criterion for the existence of the cluster
based on the experimental data.  Both of these behaviors
are reproduced in the simulations (Fig.~\ref{McNFal:fig:simus}).  Therefore,
the simulations should give insight into the origin of the bouncing
cluster.

\begin{figure*}
\begin{center}
\includegraphics[width=.67\textwidth]{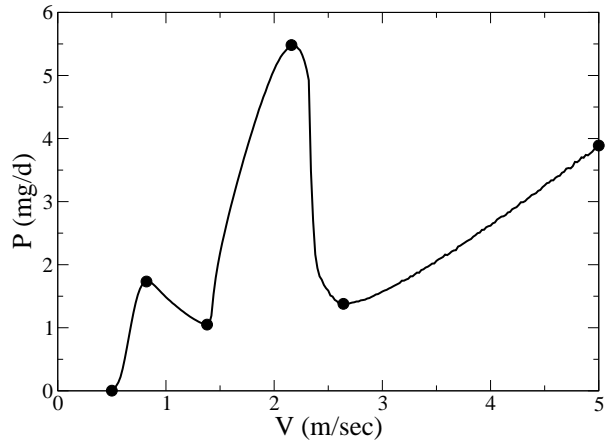}
\end{center}
\caption[]{Pressure as a function of the piston velocity, $V=4Af$, for $n=6$, 
$r=r(v)$, $h-h_0=15\;\text{mm}$ under gravity.  Simulations where done
for $f=5$ Hz to $f=50$ Hz with steps of $0.2$ Hz.
The dots mark the particular frequencies
shown in Fig.~\ref{McNFal:fig:cm}. }
\label{McNFal:fig:Presonance}
\end{figure*}

We first examine the relation between $P$ and $V$ in a case where
the bouncing cluster is observed.  One does not obtain the simple scaling in Eq.~(\ref{McNFal:eq:Pscale}) for the pressure.  Instead, one obtains a much more complicated relation between $P$ and $V$.  Indeed, as we show in
Fig.~\ref{McNFal:fig:Presonance}, the pressure displays two maxima as a function of $V$.
These maxima correspond to the bouncing cluster shown in Figs~\ref{McNFal:fig:photos}b
and \ref{McNFal:fig:simus}b.  The bouncing cluster transports momentum very effectively
between the piston and the wall.  When the bouncing cluster disappears
at $n<3$, the maxima also disappear.  These maxima make
it impossible to describe the relation between the pressure and
the forcing by a simple power law.  This is the reason why there
are no points on Fig.~\ref{McNFal:fig:thetap}d for $n>3$.  

Unlike the other results of this paper, Fig.~\ref{McNFal:fig:Presonance}
changes significantly when $L$ and $N$ are changed while holding
$n$ fixed.  When $L$ and $N$ are both divided by two, the first
maximum moves to $V\approx1\;\mathrm{m/sec}$ and the second maximum
disappears.  When $L$ and $N$ are doubled, the second maximum moves
to near $V\approx1.4\;\text{m/sec}$.  Therefore, the results of
this section should not be considered as a systematic investigation
of the resonance, but rather a descriptive introduction.  The
resonance is sensitive to $L$ because it can involve horizontal
collective motions of the particles, not just vertical ones.
Indeed, for $L=40\;\mathrm{cm}$, one can observe waves similar
to those discussed in \cite{Bizon:98}.

\begin{figure*}
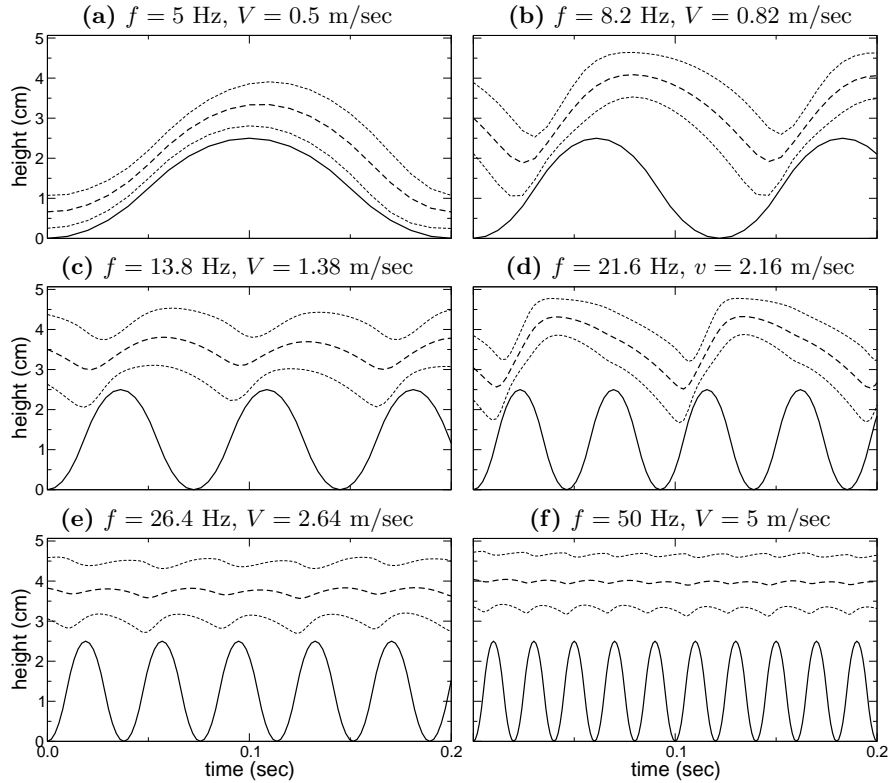

\begin{center}
\begin{tabular}{cc}
{\bf (a)} $f=5$ Hz, $V=0.5$ m/sec &
{\bf (b)} $f=8.2$ Hz, $V=0.82$ m/sec\\
\includegraphics[width=.50981\textwidth]{fig12a.eps} &
\includegraphics[width=.47019\textwidth]{fig12b.eps} \\
{\bf (c)} $f=13.8$ Hz, $V=1.38$ m/sec &
{\bf (d)} $f=21.6$ Hz, $v=2.16$ m/sec\\
\includegraphics[width=.50981\textwidth]{fig12c.eps} &
\includegraphics[width=.47019\textwidth]{fig12d.eps} \\
{\bf (e)} $f=26.4$ Hz, $V=2.64$ m/sec &
{\bf (f)} $f=50$ Hz, $V=5$ m/sec\\
\includegraphics[width=.50981\textwidth]{fig12e.eps}&
\includegraphics[width=.47019\textwidth]{fig12f.eps}\\
\end{tabular}
\end{center}
\caption{The motion of the grains under gravity at various vibration
frequencies during $0.2$ seconds,  for $n=6$, $h-h_0=25$ mm and $r=r(v)$.  
In all plots, the vertical axis is height, with the
lower boundary at the piston's lowest point, and the upper boundary
being the upper wall of the container. Solid line shows the position of the piston. The heavy dotted line shows the height 
$y_{\text{cm}}$ of the center of mass.  The upper and lower thin
dotted lines show $y_{\text{cm}} + \sigma_{\text{cm}}$ and
$y_{\text{cm}} - \sigma_{\text{cm}}$ respectively, where
$\sigma_{\text{cm}}$ is the standard deviation of the particle
heights at a given instant.  The majority of the particles are thus
contained between the two thin dotted lines.}
\label{McNFal:fig:cm}
\end{figure*}

To investigate the origins of the bouncing cluster, we show
in Fig.~\ref{McNFal:fig:cm}
the position of the piston and the approximate location of
the cluster as a function of time for various driving frequencies.
The frequencies are chosen to correspond to the maxima
and minima of the pressure, and are indicated by the large points
in Fig.~\ref{McNFal:fig:Presonance}.
At very low frequencies (Fig.~\ref{McNFal:fig:cm}a), the pressure vanishes 
because no particles hit the top of the container.  The cluster bounces
on the vibrating plate.  As the piston velocity increases, the cluster
flies higher and higher until it strikes the top.  The pressure reaches
a first maximum near $V=0.82$ m/s (Fig.~\ref{McNFal:fig:Presonance}) where there is a first resonance
between the time of flight of the cluster and the vibration period. 
These two times are such that the cluster lands on the piston just
when the piston is at its maximum upwards velocity
(Fig.~\ref{McNFal:fig:cm}b).  Note that a large number of
particles descend below the piston's maximum height.
Then as the piston rises, it sweeps up all the
particles, increasing the density of the cluster.
On the other hand, as the cluster
falls from the stationary wall, it slowly expands.

As the piston
velocity increases further, the pressure decreases, because the cluster
lands on the piston when it is near its maximum height
(Fig.~\ref{McNFal:fig:cm}c).  Then the pressure rises to a new maximum near $V=4Af=2.2$ m/s (Fig.~\ref{McNFal:fig:Presonance}) because there is a second resonance: the cluster's flight time is equal
to two piston cycles (Fig.~\ref{McNFal:fig:cm}d).  Note that the frequencies
of these two resonances ($f=8.2\;\text{Hz}$ and $f=21.6\;\text{Hz}$)
are not related by a simple factor of $2$.  The reason for this is
that the time of flight of the cluster depends in a complicated way
on the piston velocity.  Comparing Figs.~\ref{McNFal:fig:cm}b and \ref{McNFal:fig:cm}d,
one can see that the velocity of the cluster when it leaves the 
piston is close to $V$, but the time it takes the cluster to fall back
from the upper wall onto the vibrating is roughly
independent of $V$.  Note also
that the collisions between the cluster and the piston or wall takes
a finite time, as can be seen by comparing the times of the minima of
$y_{\text{cm}} + \sigma_{\text{cm}}$, $y_{\text{cm}}$, and
$y_{\text{cm}} - \sigma_{\text{cm}}$ in Fig.~\ref{McNFal:fig:cm}b.
All of these effects exclude a simple resonance
condition for the pressure maxima.  The resonant frequencies depend on
$g$ and $h-h_0$.  The resonant frequencies are proportional to $g^{1/2}$
but decrease for as $h-h_0$ is increased.  The resonant frequencies
are not sensitive to changing $n$, as long as $n$ is sufficiently large.

At higher driving
frequencies (Fig.~\ref{McNFal:fig:cm}e and Fig.~\ref{McNFal:fig:cm}f), the center
of mass of the particles changes very little over one cycle, and there
are no more resonances.  In fact, setting $g=0$ changes 
Fig.~\ref{McNFal:fig:cm}f very little.  The reason for this is that average
particle kinetic energy is much larger than $gh$, so
gravity is irrelevant.  Higher order resonances,
(where the cluster's flight time equals three or more vibration periods)
are not observed.  These resonances are probably not observed
because there is
not enough time for the particles to fall below the
piston's highest point.  Thus, the piston does not gather up all the
particles into a single
dense cluster, but rather acts to disperse the cluster.
For example, imagine that the driving
frequency is suddenly doubled in Fig.~\ref{McNFal:fig:cm}d just at the
moment when the cluster starts its descent from the wall.  When the
cluster arrives at the piston, perhaps a half of the particles
will be swept up by one vibration cycle, and a half in the next cycle.
This is how energetic forcing breaks up clusters.

\section{Conclusions}

In Sec.~\ref{McNFal:sec:experiment}, we brought simulations as close as
possible to the experiments.  To bring simulations into agreement
with the experiments, it is essential to use
a velocity dependent coefficient of restitution.  It is especially
important to take into account plastic deformations that cause
the restitution coefficient to decrease rapidly with increasing
impact velocity.  Indeed, the restitution coefficient for strongly vibrated steel spheres is very far from the constant values of $r=0.95$ or $r=0.9$ that are often cited in simulations as typical for steel spheres.  We also noted that
it is very important to take into account the number of particle
layers $n$.  The dependence of the pressure $P$ on the piston velocity $V$ changes with $n$.  It
is not accurate to speak of ``a'' scaling exponent for the pressure
in terms of $V$: this exponent depends continuously on $n$, and does not exist at high density ($n>3$) under gravity, due to the clustering instability.

In Sec.~\ref{McNFal:sec:clusters}, we investigated this question of clustering.
When there is no gravity, a loose cluster forms against the stationary wall
opposite the vibrating piston.  Due to the velocity dependent restitution coefficient, this cluster acts as a buffer for fastest particles, and leads to a simple non-extensive scaling $P\propto N^0V$.  Unlike the case with gravity,
the scaling holds for a wide range of parameters.
We discussed the origin the scaling, and remarked that it should be
observable in microgravity experiments.  We also studied the
effect of clustering in the presence of gravity.  In this case, the
clusters give a very complicated behavior, because there are
interactions between the period of vibration and the flight time
of the cluster.  Finally, if the vibration frequency is strong enough,
the cluster can be broken up, because the cluster-piston collision
time becomes longer than vibration period.

{\bf Acknowledgments}\\
We thank St\'ephan Fauve for fruitful discussions. The authors gratefully acknowledge the hospitality of the ENS-Lyon physics department which made this collaboration possible.

\bibliography{paper}

\end{document}